\documentclass[prb,superscriptaddress,twocolumn,showpacs]{revtex4}
\usepackage{indentfirst}
\usepackage{graphicx}
\usepackage{amsmath}
\usepackage{color}

\begin{document}

\title{Artificial molecular quantum rings under magnetic field influence}
\author{L. K. Castelano \footnote{Present address:
Department of Physics, University of California San Diego, La Jolla,
California 92093-0319, USA,
email:lcastelano@physics.ucsd.edu}}\author{ G.-Q. Hai}
\address{Instituto de F\'isica de S\~ao Carlos, Universidade de
S\~ao Paulo, 13560-970, S\~ao Carlos, SP, Brazil}
\author{B. Partoens} \author{F. M. Peeters}
\address{Department of Physics, University of Antwerp,
Groenenborgerlaan 171, B-2020 Antwerpen, Belgium}

\begin{abstract}
The ground states of few electrons confined in two vertically
coupled quantum rings in the presence of an external magnetic field
are studied systematically within the current spin-density
functional theory. Electron-electron interactions combined with
inter-ring tunneling affects the electronic structure and the persistent
current. For small values of the external magnetic field, we
recover the zero magnetic field molecular quantum ring ground state
configurations. Increasing the magnetic field many angular momentum,
spin, and iso-spin transitions are predicted to occur in the ground state.
We show that these transitions follow certain rules, which are governed
by the parity of the number of electrons, the single particle picture,
the Hund's rules and many-body effects.
\end{abstract}

 \maketitle

\section{Introduction}
The physics of semiconductor nanostructures has been a subject
extensively studied since the experimental realization of quantum
dots in the 1980s\cite{ekimov}. This system is very interesting due
to their similarity with atoms and the facility in controlling their
electronic, magnetic and optical properties\cite{bqds}. With the
advances in experimental techniques to fabricate
nanostructures, a novel ring-shaped nanostructure was pursued and
realized through several different approaches, \textit{e.g}.\
nano-lithography (\textit{e.g}.\ atomic force microscope
patterning)\cite{held}, droplet MBE epitaxy\cite{gong} and strain
induced self-organization\cite{garcia}. These ring-shaped
nanostructures, the so-called quantum rings (QRs), are known for their
Aharonov-Bohm effect and its persistent current\cite{AB} where
the quantum interference phenomenon leads to oscillations in the
current.

Such ring-shaped nanostructures can also be coupled in the lateral
or vertical configuration forming a ``benzene-like'' artificial molecule.
When QRs structures are coupled they open the avenue of controlling
and manipulating some fundamental quantities, as \textit{e.g.,}\ the
electron-electron
interaction\cite{lkc,malet,saiga,szaf1,szaf2,szaf3,royo}, the
integer and fractional Aharonov-Bohm oscillation\cite{hebao}, the
electron relaxation\cite{giohai} and the coupling of direct-indirect
excitons\cite{ulloa}. Quantum ring molecules (QRMs) were synthesized
experimentally using MBE technology in the form of vertically
stacked layers of self-assembled QRs\cite{granados,suarez} and
concentric double QRs\cite{mano,kuroda}. And recently, the
Aharonov-Bohm oscillation was observed in self-assembled InAs/GaAs
quantum rings containing only a single confined electron\cite{kleemans}.

In a previous work\cite{lkcpers}, we investigated theoretically the
persistent current in two vertically coupled quantum rings (CQRs)
containing 6 electrons in a wide range of inter-ring distances up to
$\sim10^4$\AA. The motivation of this work was to understand the
inter-ring quantum tunneling effects on the persistent current of
two interacting coupled rings. In order to analyze such effects, we
considered two different situations. First, we assumed that each
quantum ring (QR) contains 3 electrons and interact with each other
only via Coulomb potential. In the second situation, we included the
possibility of tunneling between electrons localized in different
QRs. We compared both situations and we found that the persistent
current is altered significantly by the quantum tunneling, which
allows the exchange interaction between electrons localized in
different QRs. Moreover, we found that an applied vertical gate can
be used to control the persistent current in such a system.

In the present work, we apply the current spin-density functional
theory (CSDFT) to determine systematically the ground states of two
vertically CQRs in the presence of an external magnetic field. The
two quantum rings are tunnel coupled and form a simple artificial
molecule. Earlier, this method was employed to investigate the
ground state properties of a single quantum ring\cite{lin,viefers}
and it was proved to be a useful tool to determine the properties of
such systems where confinement, Coulomb interaction, spin
polarization, and magnetic field are present at the same time. In
the two CQRs case, the inter-ring coupling plays an important role
and by varying the inter-ring distance between the QRs and the
applied magnetic field we found a rich variety of ground state
configurations for the systems of $N$=3,4,5, and 6 electrons. Such
results are compiled in ``phases diagrams''. Increasing the
interring distance, the isospin quantum number decreases
monotonously and the interring exchange-correlation interaction
plays an essential role forming new molecular states. Also, we found
some rules based on the single-particle picture to estimate the
ground state configuration, which might be used as a guide to
experimental analysis in the future. The persistent current as a
function of either magnetic field or distance between the QRs is
determined for a different number of confined electrons $N$=3,4,5,
and 6, thereby completing our previous work\cite{lkcpers}. We verify
that the ground state configuration and the persistent current are
inter-twined. Therefore, through measurements of the persistent
current in CQRs, the ground state configuration can be accessed
experimentally. Furthermore, we evaluate the persistent
spin-current, which is given by the difference between the spin-up
current and spin-down current. We verify that the persistent
spin-current is closely related to the ground state configuration of
the two CQRs as well.

The present paper is organized as follows. In Sect. II we present
our theoretical model within the framework of CSDFT. In Sect. III we
study the ground state properties of the CQR molecules. The phase
diagrams of the ground state configurations of few electron quantum
ring molecules in external magnetic field are obtained. In Sect. IV,
we discuss the persistent current in the CQRs and we conclude our
work in Sect. V.

\section{Theoretical model}
Within the current spin-density functional theory \cite{vignale}, we
study the magnetic field dependence of the ground state of two
vertically coupled GaAs quantum rings containing few electrons. The lateral
electron confinement in the CQRs are
described by the displaced parabolic potential model
$V(r)=\frac{1}{2}m^{\ast }\omega _{0}^{2}(r-r_{0})^{2}$ in the
$xy$-plane, where $\mathbf{r}=(x,y)=(r,\theta ) $, $\omega _{0}$ is
the confinement frequency and $r_{0}$ is the radius of the ring. The
two stacked identical rings are coupled in the $z$ direction with
the potential $V(z)$ described by two coupled symmetric quantum
wells with a barrier of finite height. The quantum wells are assumed
to be $W=120$ \AA $\;$ wide with the height of the barrier $V_{0}=250$ meV between them.
For these parameters we find\cite{lkc} the following expression for the energy
splitting $\Delta =22.86$ $\exp [-d($\AA\ $)/13.455]$ meV between
the two lowest levels in the coupled quantum wells separated by a
distance $d$. We also consider a homogeneous magnetic
field $\mathbf{B}=B\mathbf{e}_z$ applied perpendicularly to the
$xy$-plane, which is described by the vector potential
$\mathbf{A}=Br\mathbf{e}_{\theta}/2$ taken in the symmetric gauge.
The Kohn-Sham orbitals $\psi _{jnm\sigma }(\mathbf{r})=\exp
(-im\theta )\phi _{nm\sigma }(r)Z_j(z)$ are used to express the
density and ground state energy.
The Kohn-Sham equation in CSDFT for the CQRs is given by%
\begin{eqnarray}
\left[ -\frac{\hbar ^{2}}{2m^{\ast }}\left( \frac{\partial
^{2}}{\partial r^{2}}+\frac{1}{r}\frac{\partial }{\partial
r}-\frac{m^{2}}{r^{2}}+\frac{
\partial ^{2}}{\partial z^{2}}\right)
+\frac{1}{2}m^{\ast }\omega
_{0}^{2}(r-r_{0})^{2}\right].\nonumber \\
 \left. -\frac{m\hbar\omega_{c}}{2}+\frac{1}{8}m^{\ast }\omega
_{c}^{2}r^{2}-\frac{me\hbar}{c m^\ast}\frac{\mathbf{A}_{xc}}{r}+
V_{xc,\sigma }(r)
+V(z)+\right.\nonumber \\
 \left.V_{H}^{\rm{intra}}(r)+V_{H}^{\rm{inter}}(r)\right]
 \phi_{nm\sigma }(r)Z_j(z)
 =\epsilon^j _{nm\sigma }\phi _{nm\sigma }(r)Z_j(z)\label{kseq},
\end{eqnarray}
where $\sigma =\uparrow $ or $\downarrow $ is the $z$ component of
the electron spin and $\omega_c=eB/cm^\ast$ is the cyclotron
frequency. The total density in the rings is $\rho (r)=\sum_{\sigma
}\sum_{n,m}^{N_{\sigma }}|\phi _{nm\sigma }(r)|^{2}$. Because we are
adopting two identical rings, the density in each ring is half this
total density. In the calculation, we approximate the density in the
$z$ direction by $\delta$-functions at the center of the quantum
wells. This approximation has been used in our previous
work\cite{lkc} for two coupled quantum rings (QRs) with $B=0$ and
will not change our results qualitatively. The intra-ring and
inter-ring Hartree potentials are given by
\begin{equation}
V_{H}^{\rm{intra}}(r)=\int d\mathbf{r^{\prime }}\frac{e^{2}\rho
(r^{\prime })/2}{\varepsilon |\mathbf{r}-\mathbf{r^{\prime }}|},
\end{equation}
and
\begin{equation}
V_{H}^{\rm{inter}}(r)=\int d\mathbf{r^{\prime }}\frac{e^{2}\rho
(r^{\prime })/2}{\varepsilon |\mathbf{r}-\mathbf{r^{\prime
}}+\mathbf{d}|},
\end{equation}
respectively, with the inter-ring distance $d=|\mathbf{d}|$.

In CSDFT all the quantities are functionals depending on the spin-up
($\rho^\uparrow$) and spin-down ($\rho^\downarrow$) densities, and
the vorticity
$\mathbf{\mathcal{V}}(r)=\frac{cm^\ast}{e}\nabla\times\frac{\mathbf{j}_p(r)}{\rho(r)}$.
Therefore the exchange-correlation scalar and vector potential
can also be written as a functional of these quantities, which are given
respectively by
\begin{eqnarray}\label{exchan}
V_{xc,\sigma}(r)=\left.\frac{\partial E_{xc}
[\frac{\rho\uparrow}{^2},\frac{\rho\downarrow}{^2},\mathcal{V}]}{\partial
\rho_\sigma}
\right|_{\rho_{-\sigma},\mathcal{V}}-\frac{e}{c}A_{xc}\frac{j_p(r)}{\rho(r)},\nonumber\\
\frac{e}{c}\mathbf{A}_{xc}(r)=-\mathbf{e}_\theta\left.\frac{2cm^\ast}{2\rho(r)}\frac{\partial
}{\partial r}\left(\frac{\partial E_{xc}
[\frac{\rho\uparrow}{^2},\frac{\rho\downarrow}{^2},\mathbf{\mathcal{V}}]}{\partial
 \mathbf{\mathcal{V}}}\right)\right|_{\rho,\xi}.
\end{eqnarray}
where $\mathbf{j}_p (r)$ is the paramagnetic current. Also in CSDFT,
an expression for the exchange-correlation energy functional must be
found and here we adopt the local density approximation remembering
that in the bulk the total current density is zero
($\mathbf{\mathcal{V}}=\mathbf{B}$). Thus the local density
approximation can be implemented, which reads $E_{xc}= \int
d\mathbf{r} \frac{\rho(r)}{2}\epsilon _{xc}[\rho^{\uparrow },\rho
^{\downarrow },\mathbf{\mathcal{V}}]$,
where $\epsilon _{xc}$ is the exchange-correlation energy per particle
of the uniform two-dimensional electron gas in
a magnetic field $\mathbf{\mathcal{V}}=\mathbf{B}$. Following
Ref.~\cite{bart} we assume that $\epsilon _{xc}[\rho^{\uparrow },\rho
^{\downarrow },|\mathbf{\mathcal{V}}|=B]=\left(\epsilon^{LWM}
_{xc}[\rho ,\nu]+\nu^4\epsilon^{TC} _{xc}[\rho ^{\uparrow },\rho
^{\downarrow }]\right)/(1+\nu^4)$, where $\nu=2\pi\hbar c\rho/eB$ is
the filling factor. The expression for $\epsilon _{xc}$ connects the
fitted form of Levesque \emph{et al.} \cite{lesb} $\epsilon^{LWM}
_{xc}[\rho ,\nu]$, which is valid for large magnetic field, to the
form given by Tanatar-Ceperley \cite{tan} $\epsilon^{TC} _{xc}[\rho
^{\uparrow },\rho ^{\downarrow }]$ valid for zero magnetic field. We
expand the eigenfunctions $\phi _{nm\sigma }(r)$ in the well-known
Fock-Darwin basis in order to solve the Kohn-Sham equation.
\begin{figure}[t]
\centering
\includegraphics[angle=0,scale=.825]{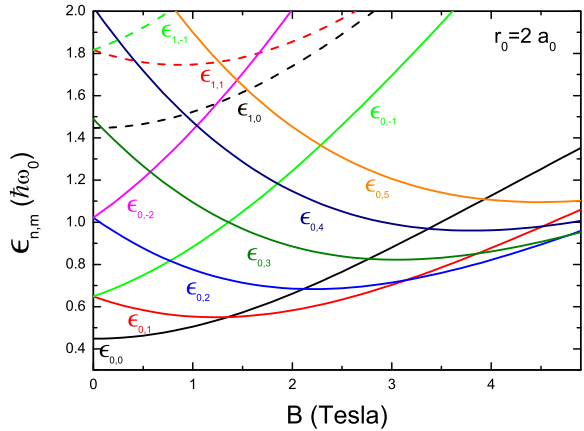} %

\caption{(Color online) Electron energy levels
$\protect\epsilon_{n,m}$ of a single quantum ring as a function of
the magnetic field. The ring radius is fixed at $r_0=2a_0$.}
\end{figure}
\begin{figure*}[htbp]
\begin{center}$
  \begin{array}{cc}
  \includegraphics[angle=0,scale=.8]{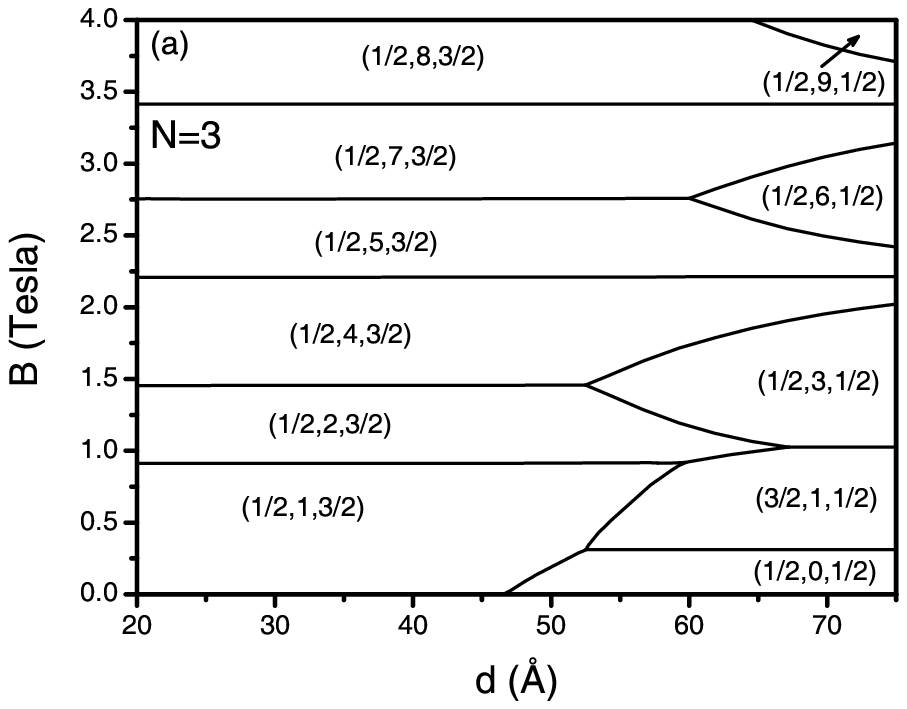} &  \includegraphics[angle=0,scale=.8]{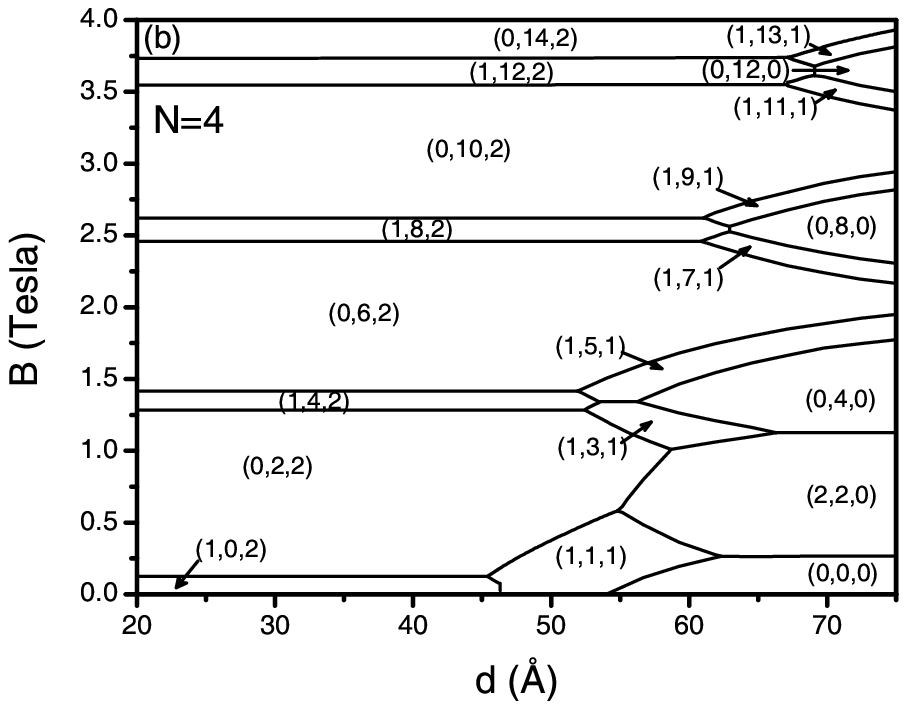}\\
  \includegraphics[angle=0,scale=.8]{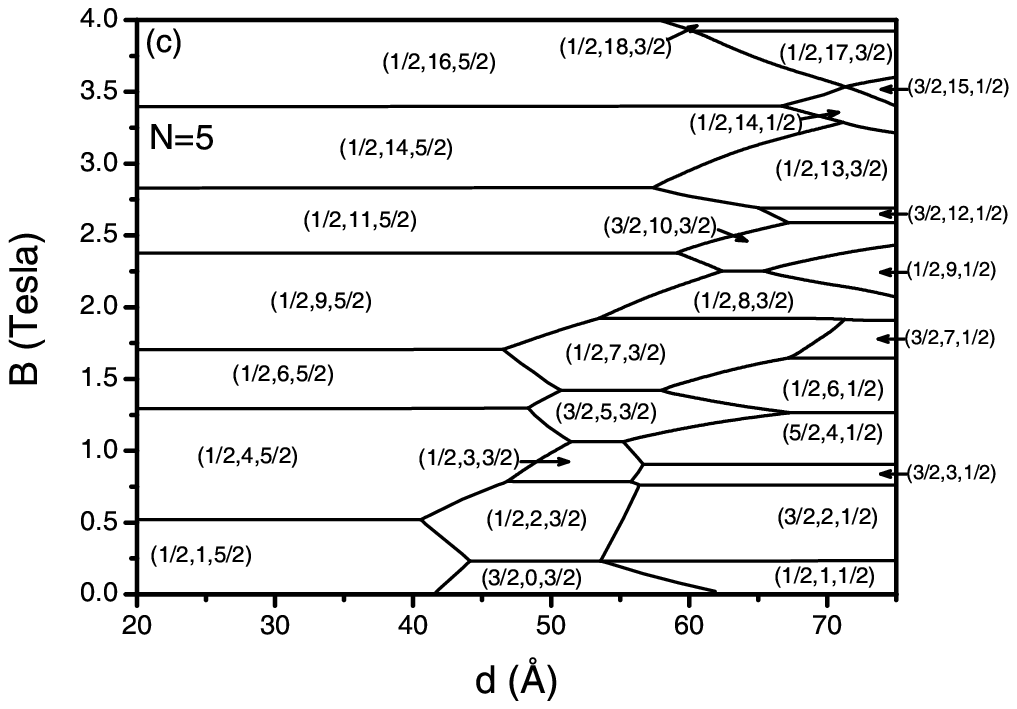} & \includegraphics[angle=0,scale=.8]{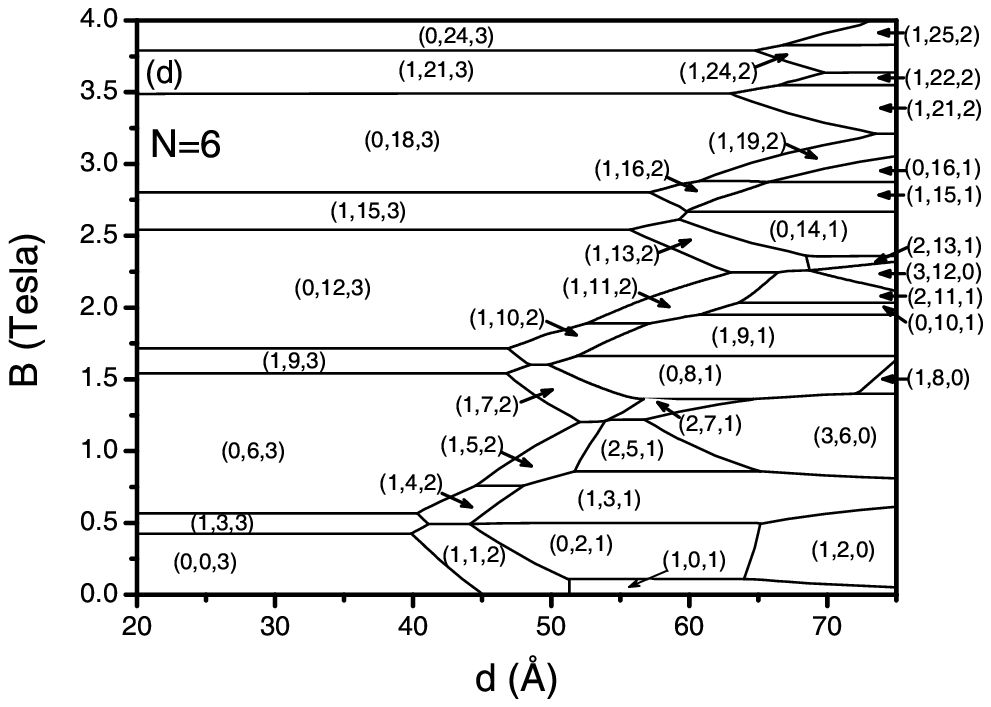}
    \end{array}$
    \caption{The phase diagram of the ground state of CQRs
for a fixed ring radius ($r_0=2a_0$) for: (a) $N=$ 3, (b) $N=$ 4,
(c) $N=$ 5, and (d) $N=$ 6. The insets indicate the three quantum
numbers $(S_{z},M_{z},I_{z})$ which designate a ground state
configuration.}
\end{center}

\end{figure*}

The ground state (GS) energy of the coupled rings as a function of
magnetic field is obtained from
\begin{eqnarray}
E(N)=\sum_{jnm\sigma }\epsilon^j _{nm\sigma }-\frac{1}{2}\int
d\mathbf{r}\rho
(r)\left[ V_{H}^{\rm{intra}}(r)+V_{H}^{\rm{inter}}(r)\right]  \nonumber \\
-\sum_{\sigma }\int d\mathbf{r}\rho ^{\sigma }(r)V_{xc,\sigma}(r)
-\frac{e}{c}\int d\mathbf{r}j_p(r)A_{xc}(r)\nonumber \\
+2E_{xc}\left[\rho ^{\uparrow }/2,\rho ^{\downarrow
}/2,\mathbf{\mathcal{V}}\right],
\end{eqnarray}%
and the paramagnetic current density is given by
\begin{equation}\label{parcur}
 \mathbf{j}_p (r)=-\frac{\hbar}{m^\ast r}\mathbf{e}_\theta\sum_{\sigma
}\sum_{n,m}^{N_{\sigma }}m|\phi _{nm\sigma }(r)|^{2}.
\end{equation}
The measurable current density is equal to
$\mathbf{j}(r)=\mathbf{j}_p(r)+(e/cm^\ast)\rho(r)\mathbf{A}(r)$,
where the second part corresponds to the diamagnetic current
density.

In the $z$ direction we consider only the two lowest levels of the
quantum wells that connect the two quantum rings. They are the
symmetric bonding level and the antisymmetric antibonding level. The
contribution from excited states due to confinement in the $z$
direction is neglected because the confinement in the $z$ direction
is much stronger than that in the plane. Therefore the motion in the
$z$ direction may be assumed to be decoupled from the in-plane
motion and the Kohn-Sham equations can be solved separately from
the Schr\"odinger equation in the $z$ direction that describes the
tunneling between the quantum rings.
In the limits of small and large inter-ring distance $d$, the results for
single quantum rings are recovered. On the other hand, in the limit
of small ring radius ($r_{0}\rightarrow 0$), results of the CQDs are
recovered\cite{bart}, too.

\section{Ground state configurations}

The energy levels $\epsilon_{n,m}$ of a single quantum ring with
fixed radius $r_0=$2 $a_0$ ($r_0=$300 \AA) are shown in Fig.~1 as a
function of the magnetic field, where
$a_0=\sqrt{\hbar/m^*\omega_0}$. The energy levels are labeled by the
radial quantum number $n=0,1,2,...,$ and the angular quantum number
$m=0,\pm 1,\pm 2,...$. The applied magnetic field breaks the $\pm m$
degeneracy and leads to angular momentum transitions in the ground
state. Although we assume a fixed value for the ring radius $r_0=$2
$a_0$, the effects observed in Fig.~1 remains the same for different
ring radius. The only difference is that the value of the magnetic
field where the crossing occurs is rescaled and assumes a smaller
value when the ring radius is increased. Using the single-particle
picture one expects already that the few electron ground state of
the CQRs will be strongly affected by the magnetic field. In
contrast, qualitative different behavior from two vertically coupled
quantum dots was found only when $N>8$ in the absence of the
magnetic field \cite{lkc}.

For the CQRs, the ground state configuration changes as a function
of the inter-ring distance (tunneling energy) and magnetic field.
The phase diagrams presented in Fig.~2 show the different
configurations of the ground state of CQRs for a fixed ring radius
$r_0$=2 $a_0$ and confinement energy $\hbar\omega_0=$5 meV as a
function of the magnetic field and the inter-ring distance for (a)
$N=$ 3, (b) $N=$ 4, (c) $N=$ 5 and (d) $N=$ 6. To perform the
numerical calculation, we use typical GaAs values for the effective
mass $m^*=0.067m_0$ and the dielectric constant $\epsilon_s=$12.4.
For small magnetic field the GS configurations are the same as found
previously for zero magnetic field \cite{lkc}. With increasing
magnetic field, many transitions in the ground state are observed,
as can be seen in Figs.~2(a-d). The ground state phases are labeled
by three quantum numbers $(S_{z},M_{z},I_{z})$: total spin $S_{z}$,
total angular momentum $M_{z}$ and the isospin quantum number
$I_{z}$, which is the difference between the number of electrons in
the bonding state and in the antibonding state divided by 2.

For large inter-ring distance the two rings become decoupled. On the
other hand, when the distance between them is small, they are
strongly coupled acting as a single one with isospin number
$I_z=N/2$ where all the electrons are situated in the bonding state.
In the latter regime, six different ground state configurations with
$I_z=3/2$ are found for $N$=3 ($B<4$ T) in Fig.~2(a). Qualitatively
the different GS configurations as a function of the magnetic field
can be understood through the single-particle (SP) picture in Fig.~1
together with Hund's rules. For example, when $B<1.37$ T the SP
energy level $\epsilon_{0,0}$ is lower than $\epsilon_{0,1}$, hence
it can be filled with two electrons with opposite spin and the third
electron occupies the state $\epsilon_{0,1}$ with spin up, resulting
in the GS configuration $(1/2,1,3/2)$. But when $B>1.37$ T, the
energy $\epsilon_{0,1}$ is lower than $\epsilon_{0,0}$ and it
becomes energetically more favorable to fill the state
$\epsilon_{0,1}$ with two electrons instead of $\epsilon_{0,0}$ and
the transition $(1/2,1,3/2)\rightarrow(1/2,2,3/2)$ in the GS takes
place. We also notice that in Fig.~2(a) this transition occurs
effectively at lower magnetic field $B=0.85$ T because of the
many-body effects. The other transitions
$(1/2,2,3/2)\rightarrow(1/2,4,3/2)\rightarrow(1/2,5,3/2)\rightarrow(1/2,7,3/2)\rightarrow(1/2,8,3/2)$
are due to the same mechanism just explained. When the inter-ring
distance increases, the difference between the bonding and
antibonding states decreases and transitions in the GS
configurations can be observed too. For large inter-ring distance in
Fig.~2(a), the GS configurations with one electron in the
antibonding state ($I_z=1/2$) are found. When the
bonding-antibonding energy splitting $\Delta$ is less than the
difference between two subsequent lateral bonding states the
electron changes to the lowest unoccupied antibonding state yielding
a GS transition. The state $(3/2,1,1/2)$ can not be explained only
using the SP picture, because it has three aligned spins in
different SP states. This is a clear manifestation of the many-body
effects, where the total energy is minimized by the exchange
interaction, when two electrons are in the same quantum state.

For $N$=4 in Fig.~2(b) we observe eight different GS configurations
for $B<4$ T in the strong coupling regime ($I_z=2$) consistent with
the SP picture and Hund's rule. The total spin in the $z$ direction
of these GS configurations alternates between 0 and 1 as a function
of the external magnetic field. Increasing the inter-ring distance
we found GS configurations with one ($I_z=1$) and two ($I_z=0$)
electrons occupying the antibonding sates. Notice that for a fixed
isospin $I_z=$1 or 2, the total angular momentum changes by $N/2=$2
as function of the magnetic field. This fact is related to the even
number of electrons ($N=$4) occupying the CQRs. In the strong
coupling regime, the even number of electrons can be arranged in
such a way that $S_z$ oscillates between 0 and 1 and $M_z$ changes
by $N/2$ with increasing the applied magnetic field. Also the state
$(2,2,0)$ has all spins aligned due to the exchange interaction.

For $N$=5 seven different GS configurations in the strong coupling
regime ($I_z=5/2$) are found in Fig.~2(c) for $B<4$ T. All of them
are of total spin $S_z=1/2$. In this regime with increasing magnetic
field, the total angular momentum alternately changes by 3
[$(1/2,1,5/2)\rightarrow (1/2,4,5/2)$] and by 2
[$(1/2,4,5/2)\rightarrow (1/2,6,5/2)$]. Since the number of
electrons is $N=$5, two SP-levels are filled by two pair electrons
and the third level by a single electron, which causes a change of
the angular momentum by either 2 or 3 with increasing magnetic
field. Increasing the inter-ring distance we found GS configurations
with one ($I_z=3/2$) and two ($I_z=1/2$) electrons occupying the
antibonding states. For these configurations the total spin can
achieve values larger than 1/2. The states with total spin $S_z=3/2$
and isospin $I_z=3/2$ or $I_z=1/2$ are induced by the exchange
interaction. Also the exchange interaction leads to the emergence of
the maximum spin polarized state $(5/2,4,1/2)$. For intermediate
$d$-values, we have $\Delta M_z$=1 transitions with increasing
magnetic field corresponding to a single angular momentum increase
of a single electron in this molecular state.

\begin{figure}[t]
\centering
\includegraphics[angle=0,scale=.825]{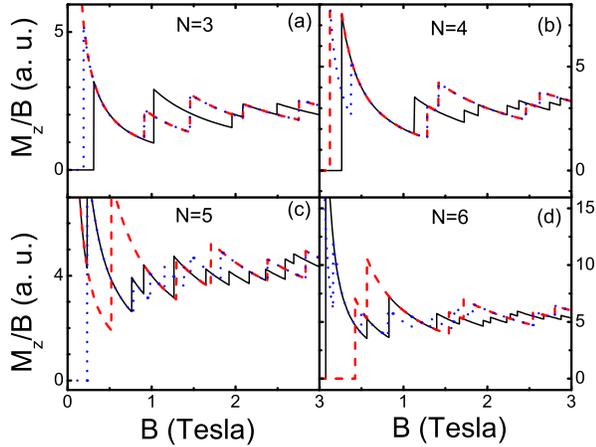} %
\caption{ (Color online) $M_z/B$ as a function of magnetic field for
different inter-ring separations $d=$30 \AA\ (red dash curves), 50
\AA\ (blue dotted curves), and 70 \AA\ (black solid curves).}
\end{figure}
\begin{figure}[b]
\centering
\includegraphics[angle=0,scale=.825]{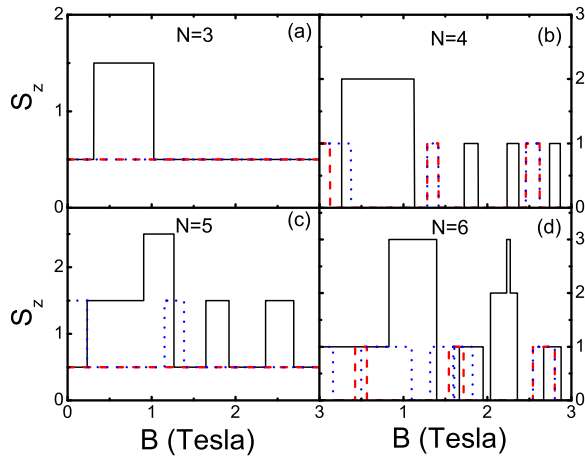} %
\caption{(Color online) The same as Fig.~3 but now for spin $S_z$.}
\end{figure}

\begin{figure}[t]
\centering
\includegraphics[angle=0,scale=.825]{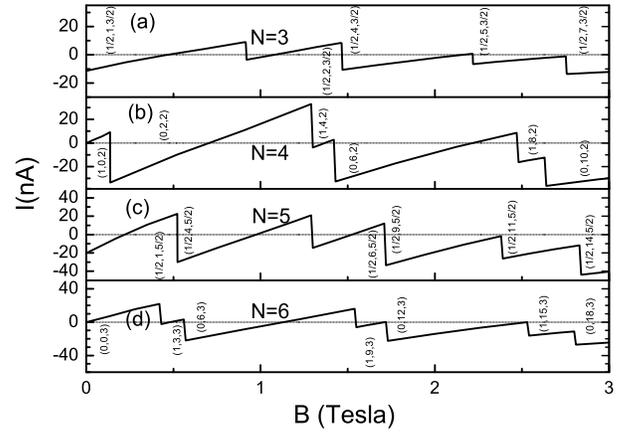} %
\caption {The persistent current as a function of the magnetic field
in the strong coupling regime ($d=30$ \AA) for CQRs with different
number of electrons: (a) $N=3$, (b) $N=4$, (c) $N=5$, and (d) $N=6$.}

\end{figure}

For $N=6$ in Fig.~2(d) nine different GS configurations are found in
the strong coupling regime ($I_z$=3) for $B<4$ T. Notice that for
$I_z$=3 the total angular momentum changes by $N/2$=3 as function of
the magnetic field due to the even number of electrons ($N$=6) in
the CQRs. Increasing the inter-ring distance we found the GS
configurations with one ($I_z=2$), two ($I_z=1$) and three ($I_z=0$)
electrons occupying the antibonding states. When $I_z=2$, the total
spin is $S_z=1$ because the bonding states are occupied by 5
electrons and the GS configurations are the same as found earlier
for $N=5$ in the strong coupling regime, which always have one
electron unpaired and the lowest antibonding state is filled with
one more spin 1/2 electron. Many different states are induced by the
exchange interaction when $N=6$, \textit{e.g.,}, the eight states
composed with total spin $S_z=1$ or $S_z=2$ and isospin $I_z=1$. The
GS configurations $(3,6,0)$ and $(3,12,0)$ have all spins aligned.

In phase $(3,6,0)$, 3 electrons both in the bonding and the
antibonding states occupy successive angular momentum states and the
total angular momentum $M_z=\frac{N}{2}(\frac{N}{2}-1)$ which is the
densest spin polarized electron configuration available in a quantum
ring molecule of 6 electrons. Such a state is referred to as the
maximum density droplet and was observed experimentally in a quantum
dot in the presence of a magnetic field.\cite{mdd} However, we
notice that there is no corresponding single quantum ring phase for
such a configuration. It exists only in the CQRs because of the
many-body interactions combined with the inter-ring quantum
tunneling effect. In fact, the phases (5/2,4,1/2) for $N=5$ found in
Fig.~2(c), (2,2,0) for $N=4$ in Fig.~2(b), and (3/2,1,1/2) for $N=3$
in Fig.~2(a) are the maximum density droplet states in the QR
molecules.
\begin{figure}[t]
\centering
\includegraphics[angle=0,scale=.825]{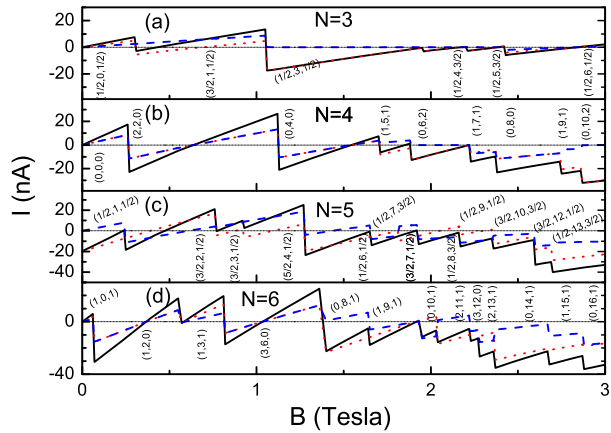} %
\caption{(Color online) The persistent current (solid curves) as a
function of the magnetic field in the weak coupling regime ($d=70$
\AA) for CQRs with different number of electrons: (a) $N=$ 3, (b)
$N=$ 4, (c) $N=$ 5, and (d) $N=$ 6. The dashed (red) and dotted
(blue) curves correspond to the bonding and anti-bonding currents,
respectively.}

\end{figure}

\begin{figure}[b]
\centering
\includegraphics[angle=0,scale=.825]{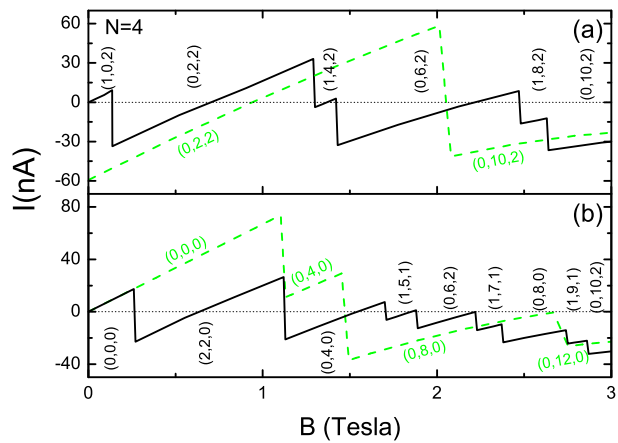}
\caption{(Color online) The persistent current obtained through CSDFT
calculation (solid black curve) and the single particle
approximation (dashed green curve), considering the CQRs with four
electrons in both (a) strong ($d=30$ \AA) and (b) weak coupling
regime ($d=70$ \AA).}

\end{figure}

\begin{figure}[t]
\centering
\includegraphics[angle=0,scale=.825]{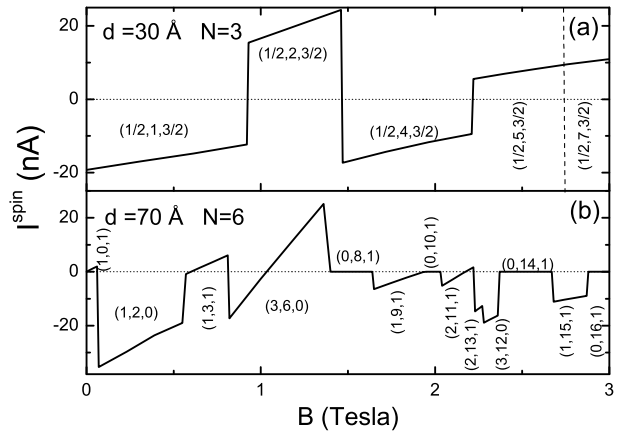} %
\caption{ The total spin-current considering (a) $N=3$ in the strong
coupling regime ($d=30$ \AA) and (b) for $N=6$ in the weak coupling
regime ($d=70$ \AA).
}
\end{figure}

In order to show more clearly the behavior of the total angular momentum
and total spin in the CQRs, we plot $M_z/B$ and $S_z$ as a function of magnetic field
in Fig. 3 and Fig. 4, respectively, for different inter-ring distances
$d=$30 \AA\ (the dash curves), 50 \AA\ (the dotted curves), and 70 \AA\
(the solid curves). The oscillation of $M_z/B$ in Fig.~3 is related to the magnetization
of the CQRs and the corresponding persistent current which will be shown below.
In Fig.~4 we see that with increasing the distance between the two rings the spin
polarization of the system is enhanced. Generally, the QR molecule in the weak coupling
regime ($d=$70 \AA) is spin-polarized in a wider range of magnetic field than
that in the strong coupling regime ($d=$30 \AA, which is practically in the atomic regime).
The maximum spin polarization $S_z=N/2$ always occurs in the molecular phases
in the weak coupling regime at finite magnetic field.

\section{Persistent current}

In Fig.~5 the persistent current is shown, determined by\cite{emperador}
$I=\int j(r) dr$, as a function of the magnetic field in
the strong coupling regime ($d=30$ \AA ) for different number of
electrons $N=$ 3, 4, 5 and 6. The diamagnetic contribution to the
persistent current in our model can be evaluated analytically and
is given by $I_d=\frac{eNB}{4\pi m^\ast c }$. When the number of electrons
in the CQRs is odd ($N$=3, 5), the persistent current oscillates
with linear segments, and the different segments appear because of the
change in the total angular momentum. As can be viewed in Figs.~3(a)
and 3(c), for $d=30$ \AA\ the total angular momentum changes as
function of the magnetic field. The total momentum increases from
$M_z^{<}$ to $M_z^{>}$ with increasing magnetic field following
the rule $M_z^{>}=M_z^{<}+(N\pm 1)/2$ indicated in Fig.~2. These
relations for the total angular momentum in the strong coupling
regime can be explained by the breaking of the $\pm m$ degeneracy
and the crossing between states with angular quantum number
$(m+1)$ and $\pm m$. For example, in the case of $N=5$ and $B<0.8$ T
the SP energy levels can be filled in the following way: two
electrons in the state $\epsilon_{0,0}$, two in
$\epsilon_{0,1}$ and one in $\epsilon_{0,-1}$, which gives the GS
$(1/2,1,5/2)$. Increasing the magnetic field the state
$\epsilon_{0,2}$ crosses the state $\epsilon_{0,-1}$ at $B=0.8$ T and
now the configuration of the GS is $(1/2,4,5/2)$ with: two
electrons in the state $\epsilon_{0,0}$, two in
$\epsilon_{0,1}$ and one in $\epsilon_{0,2}$. For $B>1.37$ T the
level $\epsilon_{0,2}$ crosses the $\epsilon_{0,1}$ and the
configurations of the GS is $(1/2,6,5/2)$ with: one electron in
the state $\epsilon_{0,0}$, two in $\epsilon_{0,1}$ and two in
$\epsilon_{0,2}$. Therefore the total angular momentum jumps from
$M_z=1$ to $M_z=4$ and from $M_z=4$ to $M_z=6$ respecting the
crossing between the SP energy levels. However, the value of the magnetic
field at which the GS transition occurs is not exactly the same as
the SP crossings because it is reduced by the electron-electron
interaction.
\begin{figure}[t]
\centering
\includegraphics[angle=0,scale=.825]{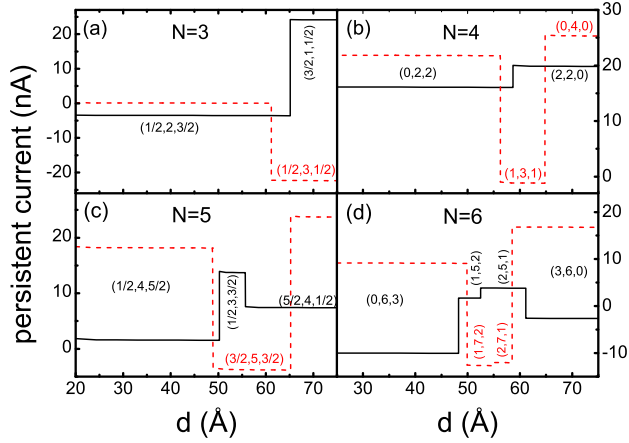} %
\caption{(Color online) The persistent current for CQRs with: (a)
$N=$ 3, (b) $N=$ 4, (c) $N=$ 5, and (d) $N=$ 6, as a function of the
distance between the two coupled quantum rings. Also, we consider
two different values of magnetic field $B=1$ T (black solid curve)
and $B=1.1$ T ($B=1.25$ T) (red dashed curve) in the upper (bottom)
panel.
}
\end{figure}

\begin{figure}[b]
\centering
\includegraphics[angle=0,scale=.825]{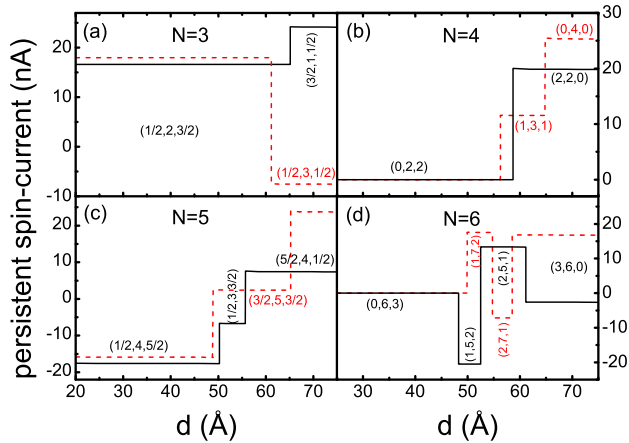} %
\caption{(Color online) The persistent spin-current for CQRs with:
(a) $N=$ 3, (b) $N=$ 4, (c) $N=$ 5, and (d) $N=$ 6, as a function of
the distance between the two coupled quantum rings. Also, we
consider two different values of magnetic field $B=1$ T (black solid
curve) and $B=1.1$ T ($B=1.25$ T) (red dashed curve) in the upper
(bottom) panel.
}
\end{figure}

When the number of electrons is even $N=4$ [see Fig.~5(b)] or $N=6$
[see Fig.~5(d)] in the strong coupling regime ($d=30$ \AA), the
persistent current also oscillates with linear segments, but an
additional fine structure is found. In this case, the difference
between two consecutive total angular momentum transitions as a
function of the magnetic field is always equal to $M_z^{>}-
M_z^{<}=N/2$ as shown in Figs.~5(b) and 5(d). When the total spin is
different from zero ($S_z\neq0$), the ground state is unstable
because it corresponds to an open-shell system, and consequently an
intermediate structure in the total persistent current appears. For
$N$-even, $S_z$ oscillates between 0 and 1 [see the dash curves in
Figs.~4(b) and 4(d)] and these transitions are reflected in the
total persistent current. The more stable linear segments correspond
to closed-shell configurations ($S_z=0$).

The persistent current as a function of the magnetic field in the
weak coupling regime ($d=70$ \AA) for different number of electrons
$N=$ 3, 4, 5 and 6 are shown in Figs. 6(a-d), respectively.
The dotted and dashed curves correspond to the bonding and
antibonding currents, respectively. The bonding (antibonding)
current is the contribution of the electrons in the bonding
(antibonding) state to the persistent current. For example,
the bonding paramagnetic
current density is found by considering only the contribution of the
bonding states in Eq.~(\ref{parcur}), and so on. The solid curve
represents the total current, which is the sum of the bonding and
antibonding currents. In the weak coupling regime, the bonding and
antibonding states can be occupied simultaneously as can be viewed
through the changes in $I_z$ indicated in Figs.~2(a-d). Therefore,
different combinations of the total angular momentum are possible as
a function of the magnetic field and any change of $M_z$ leads to a
jump in the persistent current. For large values of the magnetic field,
such jumps become more frequent because of a rapid variation of
the angular momentum at large magnetic fields and consequently
the amplitude of the oscillation in the the persistent current is reduced.

In order to understand the relevance of the many-body effects, we
show in Fig.~7 the persistent current calculated using CSDFT (solid
curve) and using the single particle approximation (dashed curve),
considering the CQRs with four electrons in both the strong [see
Fig.~7(a)] and the weak [see Fig.~7(b)] coupling regime. When the
single-particle approximation is considered, we note that most of
the little jumps in the persistent current are missing and in some
regions the single-particle results give an opposite persistent
current compared to the persistent current evaluated by CSDFT.
Therefore, through this comparison in Fig.~7 we clearly see the
importance of the many-body interactions when estimating the persistent
currents in CQRs.

We present in Fig.~8 the persistent spin-current for a CQRs with (a)
$N=3$ in the strong coupling regime ($d=30$ \AA) and (b) $N=6$ in the
weak coupling regime ($d=70$ \AA). The spin-current is defined as the
difference between the current of electrons with spin-up and with
spin-down. The paramagnetic spin-current density is given explicitly
by $\mathbf{j}^s_p (r)=-\frac{\hbar}{m^\ast
r}\mathbf{e}_\theta\left(\sum_{n,m}^{N_{\uparrow}}m|\phi
_{nm\uparrow}(r)|^{2}-\sum_{n,m}^{N_{\downarrow}}m|\phi
_{nm\downarrow}(r)|^{2}\right)$. The diamagnetic contribution for
the persistent spin-current is given by
$I^s_d=\frac{eB(N^\uparrow-N^\downarrow)}{4\pi m^\ast c }$. The
persistent spin-current is zero in the magnetic field ranges where
the total spin $S_z$ in a certain GS configuration is zero as can
be viewed in Fig.~8(b). When the CQRs is occupied by an odd number of
electrons, the persistent spin-current is never zero [see Fig.~8(a)]
because now always a not-aligned electron is present.

In addition to the external magnetic field, the distance between the
two quantum rings is another important parameter affecting the
persistent current and the spin-current. Fig.~9(a) shows the
persistent current for a CQRs with $N=3$ for different values of the
magnetic field (a) $B=1$ T (solid curve) and $B=1.1$ T (dashed
curve). The respective spin-currents are presented in Fig.~10(a).
For the CQRs with $N=4$ we plot the same quantities in Fig.~9(b) and
Fig.~10(b) for the magnetic fields $B=1$ T (solid curves) and
$B=1.1$ T (dashed curves). We plot the persistent current
(spin-current) in Fig.~9(c) (Fig.~10(c)) for a CQRs with $N=5$ for
the magnetic fields B=1 T (the solid curves) and for $B=1.25$ T (the
dashed curves). The persistent current in Fig.~9(d) and persistent
spin-current in Fig.~10(d) are evaluated for different values of the
magnetic field B=1 T (solid curves) and $B=1.25$ T (dashed curves)
for a CQRs with $N=6$. In Figs.~9-10 we notice that for a fixed
value of the magnetic field, the persistent current (spin-current)
as a function of the inter-ring distance is practically constant and
exhibits a jump when a ground state transition occurs. Therefore, by
varying the inter-ring distance for fixed value of the applied
magnetic field, the GS configuration of the CQRs can be determined.
For $N=3$ and $B=1$ T ($B=1.1$ T), we have the transition
$(1/2,2,3/2)\rightarrow (3/2,1,1/2)$ ($(1/2,2,3/2)\rightarrow
(1/2,3,1/2)$) as a function of the inter-ring distance. When $B=1$ T
the total spin changes from $S_z=1/2$ to $S_z=3/2$ and the total
angular momentum from $M_z=2$ to $M_z=1$. Therefore the total
angular momentum decreases and the persistent current increases [see
Fig.~9(a)], because of the reduction in the paramagnetic current
[see Eq.~(\ref{parcur})]. The increase in the total spin in the
$z$-direction causes an amplification of the persistent
spin-current, as can be viewed in Fig.~10(a). On the other hand,
when $B=1.1$ T the total spin does not change and the total angular
momentum decreases, which leads to a reduction of both the
persistent current and the spin-current.

When the CQRs are filled with 4 electrons and the applied magnetic
field is $B=1$ T, the transition $(0,2,2)\rightarrow (2,2,0)$ occurs
as a function of the inter-ring distance. Increasing the magnetic
field slightly ($B=1.1$ T), an intermediate GS configuration appears
and now we have the transitions $(0,2,2)\rightarrow
(1,3,1)\rightarrow (0,4,0)$. When $B=1$ T, the total spin changes
from $S_z=0$ to $S_z=2$ and the total angular momentum is not
affected. Thus the persistent current is almost unaltered [see
Fig.~9(b)] and the persistent spin-current changes abruptly [see
Fig.~10(b)], because all the carriers are spin polarized for
$S_z=2$. Moreover, when $B=1.1$ T the persistent current firstly
decreases (56\AA$<d<$65\AA) and afterwards increases ($d>$65\AA) due
to the increase of the angular momentum. The persistent spin-current
is zero for $d<$56\AA, because the net spin is zero in this range.
For 56\AA$<d<$65\AA\ the persistent spin-current is different from
zero and increases more for $d>$65\AA, because of full alignment of
the electron spins. For $N=5$, we choose two different magnetic
fields $B=1$ T and $B=1.25$ T which gives the following GS
transitions $(1/2,4,5/2)\rightarrow (1/2,3,3/2)\rightarrow
(5/2,4,1/2)$ and $(1/2,4,5/2)\rightarrow (3/2,5,3/2)\rightarrow
(5/2,4,1/2)$, respectively. The initial and final GS configurations
are the same for these two magnetic fields and both have the same
total angular momentum $M_z=4$, and only the GS configurations with
$I_z=3/2$ are different. This difference brings an opposite behavior
of the persistent current [see Fig.~9(c)] for the $I_z=3/2$ GS
configurations and in one case (B=1 T) it increases and the other
($B=1.25$ T) decreases, due to the difference of total angular
momentum in both cases. The latter-type behavior of the persistent
spin-current [see Fig.~10(c)] in both cases ($B=1$ T and $B=1.25$ T)
is a consequence of the increase in the spin polarization as a
function of the inter-ring distance. The ground state transitions
for the applied magnetic field $B=1$ T and $B=1.25$ T are
respectively $(0,6,3)\rightarrow(1,5,2)\rightarrow
(2,5,1)\rightarrow(3,6,0)$ and $(0,6,3)\rightarrow
(1,7,2)\rightarrow (2,7,1)\rightarrow(3,6,0)$, when the CQRs are
occupied with $N=6$ electrons. Again the initial and final GS
configurations for the magnetic fields chosen are the same and in
this case two intermediate configurations appear. When $B=1$ T
($B=1.25$ T), the persistent current [see Fig.~9(d)] increases
(decreases) when the inter-ring distance corresponds to one of the
intermediate configurations, because of the increase (decrease) in
angular momentum. The persistent spin-current is zero for the GS
configuration $(0,6,3)$ due to the spin-unpolarized electrons [see
Fig.~10(d)]. The persistent spin-current [see Fig.~10(d)] of the
intermediate GS configurations $(1,5,2)$ and $(2,7,1)$ ($(1,7,2)$
and $(2,5,1)$) decreases (increases) because the paramagnetic
spin-current is larger (smaller) than the diamagnetic spin-current.
The GS-configuration $(3,6,0)$ for $B=1$ T has a small spin-current
due to the compensation of the paramagnetic spin-current by the
diamagnetic spin-current, what does not happen for $B=1.25$ T, where
the diamagnetic spin-current is larger than the paramagnetic
spin-current.

\section{Summary}

We studied systematically the electronic structure and the
persistent current in two vertically coupled quantum rings as a
function of an applied external magnetic field and the inter-ring
distance. A rich variety of ground state configurations were found,
that are induced by changing the inter-ring distance and the applied
magnetic field. We found that these transitions are governed by some
rules, which are related to the even/odd number of electrons, the
single-particle picture, Hund's rules and many-body effects. The
found ground state configurations are summarized in phase diagrams,
which generalize the previous B=0 results\cite{lkc} to non-zero
values of the magnetic field. Moreover, the persistent current and
spin-current were evaluated for the CQRs and we showed that their
variation with magnetic field is governed by the values of the total
angular momentum and total spin. Therefore, once the GS
configuration is known the persistent current and spin-current can
be estimated. Furthermore, we provide useful information to
determine the ground state configuration through measurements of the
persistent current. Also we showed that the electron-electron
interaction strongly influences the sign and size of the persistent
current.

The simple rules and the complete set of results for $N$=3,4,5, and
6 electrons shown in the present work may not be compared to
available experimental data. The reason for that is because we
assumed a system composed of two symmetric QRs vertically coupled
containing few electrons and until now the experimental fabrication
of such an arrangement is quite difficult. Nonetheless, we believe
that our results and conclusions may be very useful to interpret
experimental data when those difficulties are overcome.

\acknowledgments This work was supported by FAPESP and CNPq (Brazil)
and by the Flemish Science Foundation (FWO-Vl) and the Belgium
Science Policy (IAP). Part of this work was supported by the EU
network of excellence: SANDiE.


\begin{thebibliography}{99}
\bibitem{ekimov}A. I. Ekimov, A. L. Efros, and A. A. Onushchenko, Solid State Commun. \textbf{56}, 921 (1985).

\bibitem{bqds} \textit{Quantum Dots}, L. Jacak, P. Hawrylak, and A. Wojs,
(Springer, Berlin, 1998); \textit{Quantum Dots}, T. Chakraborty,
(Elsevier, Amsterdam, 1999); \textit{Quantum Dot Heterostructures}
D. Bimberg, M. Grundmann and N. N. Ledentsov (Wiley, London, 2001);
\textit{Electron Transport in Quantum Dots} J. P. Bird (Kluwer
Academic Publishers, Boston, 2003).

\bibitem{held} R. Held, S. L\"uscher, T. Heinzel, K. Ensslin, and W. Wegscheider,
Appl. Phys. Lett. \textbf{75}, 1134 (1999); A. Fuhrer, S. L\"uscher,
T. Ihn, T. Heinzel, K. Ensslin, W. Wegscheider, and M. Bichler,
Nature (London) \textbf{413}, 822 (2001).

\bibitem{gong} Z. Gong, Z. C. Niu, S. S. Huang, Z. D. Fang,
B. Q. Sun, and J. B. Xia, Appl. Phys. Lett. \textbf{87}, 093116
(2005).

\bibitem{garcia} J. M. Garc\'ia, G. Medeiros-Ribeiro, K. Schmidt, T. Ngo,
J. L. Feng, A. Lorke, J. Kotthaus, and P. M. Petroff, Appl. Phys.
Lett. \textbf{71}, 2014 (1997); A. Lorke, R. J. Luyken, A. O.
Govorov, J. P. Kotthaus, J. M. Garc\'ia, and P. M. Petroff, Phys.
Rev. Lett. \textbf{84}, 2223 (2000).

\bibitem{AB} \textit{Aharonov-Bohm and other cyclic phenomena}, J. Hamilton
(Springer-Verlag, Berlin, 1997).

\bibitem{lkc} L. K. Castelano, G.-Q. Hai, B. Partoens, and F. M. Peeters, Phys. Rev. B \textbf{74},
045313 (2006); Braz. J. Phys. \textbf{36}, 936 (2006); Phys. Status
Solidi C \textbf{4}, 560 (2007).

\bibitem{malet} F. Malet, M. Barranco, E. Lipparini, R. Mayol, M. Pi, J. I. Climente, and J. Planelles,
 Phys. Rev. B \textbf{73}, 245324 (2006).

\bibitem{saiga} Y. Saiga, D. S. Hirashima, and J. Usukura, Phys. Rev. B \textbf{75},
045343 (2007).

\bibitem{szaf1}B. Szafran, S. Bednarek, and M. Dudziak, Phys. Rev. B \textbf{75}, 235323
(2007).

\bibitem{szaf2}  B. Szafran, Phys. Rev. B \textbf{77}, 235314 (2008); Phys. Rev. B \textbf{77}, 205313 (2008).

\bibitem{szaf3} T. Chwiej and B. Szafran, Phys. Rev. B \textbf{78},
245306 (2008).

\bibitem{royo} M. Royo, F. Malet, M. Barranco, M. Pi, and J. Planelles,
 Phys. Rev. B \textbf{78}, 165308 (2008).

\bibitem{hebao} Y. Z. He and C. G. Bao, Eur. Phys. J. B \textbf{62}, 465–470 (2008).

\bibitem{giohai} G. Piacente and G.-Q. Hai, J Appl. Phys. \textbf{101},
124308 (2007).

\bibitem{ulloa} L. G. G. V. Dias da Silva, J. M. Villas-B\^oas, and S. E. Ulloa, Phys. Rev. B \textbf{76},
155306 (2007).

\bibitem{suarez} F. Su\'arez, D. Granados, M. L. Dotor,
and J. M. Garc\'ia, Nanotechnology \textbf{15}, S126 (2004).

\bibitem{granados} D. Granados, J. M. Garc\'ia, T. Ben, and S. I. Molina,
Appl. Phys. Lett. \textbf{86}, 071918 (2005).

\bibitem{mano}  T. Mano, T. Kuroda, S. Sanguinetti, T. Ochiai, T. Tateno, J. Kim,
T. Noda, M. Kawabe, K. Sakoda, G. Kido, and N. Koguchi, Nano Lett.
\textbf{5}, 425 (2005).

\bibitem{kuroda} T. Kuroda, T. Mano, T. Ochiai, S.
Sanguinetti, K. Sakoda, G. Kido, and N. Koguchi, Phys. Rev. B
\textbf{72}, 205301 (2005).

\bibitem{kleemans} N. A. J. M. Kleemans, I. M. A. Bominaar-Silkens, V. M. Fomin, V. N. Gladilin, D. Granados, A. G. Taboada,
J. M. Garc\'ia, P. Offermans, U. Zeitler, P. C. M. Christianen, J.
C. Maan, J. T. Devreese, and P. M. Koenraad, Phys. Rev. Lett.
\textbf{99}, 146808 (2007).

\bibitem{lkcpers} L. K. Castelano, G.-Q. Hai, B. Partoens, and F. M. Peeters, Phys. Rev. B \textbf{78},
195315 (2008).

\bibitem{lin} J. C. Lin  and G. Y. Guo, Phys. Rev. B \textbf{65}, 035304 (2001).

\bibitem{viefers} S. Viefers, P. S. Deo, S. M. Reimann, M. Manninen, and M. Koskinen,
Phys. Rev. B \textbf{62}, 10668 (2000).

\bibitem{vignale} G. Vignale and M. Rasolt, Phys. Rev. Lett. \textbf{59}, 2360 (1987);
 Phys. Rev. B \textbf{37}, 10685 (1988).


\bibitem{bart} B. Partoens and F. M. Peeters, Europhys. Lett. {\bf 56}, 86 (2001).

\bibitem{lesb} D. Levesque, J. J. Weis and A. H. MacDonald, Phys. Rev. B {\bf 30}, 1056 (1984).
\bibitem{tan} B. Tanatar and D. M. Ceperley, Phys. Rev. B {\bf 39}, 5005 (1989).

\bibitem{mdd} T. H. Oosterkamp {\it et al.},  Phys. Rev. Lett. {\bf 82}, 2931 (1999);
A. MacDonald {\it et al.}, Aust. J. Phys. {\bf 46}, 345 (1993).


\bibitem{emperador} A. Emperador, M. Pi, M. Barranco and E. Lipparini, Phys. Rev. B \textbf{64}, 155304 (2001).


\end{thebibliography}
\end{document}